\documentclass[12pt, preprint]{aastex}

\slugcomment{published in \emph{Icarus}, \textbf{216}: 597--609 (2011) | doi: 10.1016/j.icarus.2011.10.001}

\shorttitle{Power Spectra of Jupiter's Clouds and Kinetic Energy}
\shortauthors{Choi and Showman}

\begin{document}

\title{Power Spectral Analysis of Jupiter's Clouds and Kinetic Energy from \emph{Cassini}}

\author{David S. Choi and Adam P. Showman}
\affil{Department of Planetary Sciences, The University of Arizona,
				Tucson, AZ 85721}
\email{dchoi@lpl.arizona.edu}

\begin{abstract}
We present suggestive evidence for an inverse energy cascade within Jupiter's atmosphere through a calculation of the power spectrum of its kinetic energy and its cloud patterns. Using \emph{Cassini} observations, we composed full-longitudinal mosaics of Jupiter's atmosphere at several wavelengths. We also utilized image pairs derived from these observations to generate full-longitudinal maps of wind vectors and atmospheric kinetic energy within Jupiter's troposphere. We computed power spectra of the image mosaics and kinetic energy maps using spherical harmonic analysis. Power spectra of Jupiter's cloud patterns imaged at certain wavelengths resemble theoretical spectra of two-dimensional turbulence, with power-law slopes near $-5/3$ and $-3$ at low and high wavenumbers, respectively. The slopes of the kinetic energy power spectrum are also near $-5/3$ at low wavenumbers. At high wavenumbers, however, the spectral slopes are relatively flatter than the theoretical prediction of $-3$. In addition, the image mosaic and kinetic energy power spectra differ with respect to the location of the transition in slopes. The transition in slope is near planetary wavenumber 70 for the kinetic energy spectra, but is typically above 200 for the image mosaic spectra. Our results also show the importance of calculating spectral slopes from full 2D velocity maps rather than 1D zonal mean velocity profiles, since at large wavenumbers the spectra differ significantly, though at low wavenumbers, the 1D zonal and full 2D kinetic energy spectra are practically indistinguishable. Furthermore, the difference between the image and kinetic energy spectra suggests some caution in the interpretation of power spectrum results solely from image mosaics and its significance for the underlying dynamics. Finally, we also report prominent variations in kinetic energy within the equatorial jet stream that appear to be associated with the 5 $\mu$m hotspots. Other eddies are present within the flow collar of the Great Red Spot, suggesting caution when interpreting snapshots of the flow inside these features as representative of a time-averaged state.
\end{abstract}

\keywords{Jupiter, atmosphere --- Atmospheres, dynamics --- Atmospheres, structure}

\section{Introduction}

The dominant feature of Jupiter's dynamic atmosphere is the alternating structure of tropospheric zonal jet streams that are quasi-symmetric across the equator. The wind shear of the jet streams corresponds well with an interchanging pattern of horizontal cloud bands called zones and belts, which have high and low albedo, respectively. Despite changes in the overall appearance of zones and belts over the past few decades, the jet streams have remained relatively constant during the era of solar system spacecraft exploration, with only a modest change in the speeds and shapes in a few of the 20+ observed jets \citep{Ingersoll04}. These jets possess a substantial portion of the atmospheric kinetic energy, and the energy source for their generation and maintenance against dissipation are long-standing questions.

Two main hypotheses for Jovian jet stream formation and maintenance have been advanced: shallow-forcing models, wherein eddies generated by moist convection or baroclinic instabilities (resulting from differential solar heating across latitudinal cloud bands) pump the jets, and deep-forcing models, wherein convection throughout Jupiter's outer layer of molecular hydrogen drives differential rotation within the planet's interior that manifests as jets in the troposphere \citep[for a review, see][]{Vasavada05}. Many of these theories invoke the idea that, in a quasi-two-dimensional fluid such as an atmosphere, turbulent interactions transfer energy from small length scales to large ones.  In the presence of a $\beta$ effect, this interaction can allow Jovian-style jets to emerge from small-scale forcing \citep[e.g.,][]{Williams78, Cho96, Nozawa97, Huang98, Showman07, Scott07, Sukoriansky07, Heimpel05, Lian08, Lian10, Schneider09}.  Most commonly, this upscale energy transfer is assumed to take the form of an {\it inverse energy cascade}, where strong turbulent interactions between eddies transfer kinetic energy through a continuous sequence of ever-increasing length scales and, under appropriate conditions, the kinetic energy exhibits a power-law dependence on wavenumber over some so-called {\it inertial range.}  Theoretically, there are two inertial ranges for a two-dimensional fluid energized by a forcing mechanism at some discrete length scale \citep{Vallis06}. One domain encompasses large length scales (small wavenumbers) down to the forcing scale and marks an inverse energy cascade regime, indicated by the region of the energy spectrum with power-law slopes of -5/3. The second domain comprises small length scales (large wavenumbers) up to the forcing scale and is the realm of enstrophy transfer, where enstrophy (one-half the square of relative vorticity) cascades from large to small length scales. This is marked by a region of the energy spectrum with slopes at approximately -3.

Although interaction of small-scale eddies with the $\beta$ effect is clearly crucial to Jovian jet formation, debate exists on whether the process necessarily involves an inverse energy cascade. While some authors favor an inverse-energy-cascade framework \citep{Galperin01, Sukoriansky07}, others have suggested that zonal jets can result directly from a mechanism where small eddies directly transfer their energy to the jets \citep{Huang98, Schneider06, Schneider09}. Such a wave/mean-flow interaction would not necessarily require the strong eddy-eddy interactions, and transfer of energy across a continuous progression of increasing wavelengths, implied by an inverse cascade.  

There is therefore a strong motivation to determine, observationally, whether upscale energy transfer occurs in Jupiter's atmosphere. Qualitative support for upscale energy transfer---broadly defined---is provided by observations of prevalent vortex mergers on Jupiter \citep{Maclow86, Li04, Sanchez01}. In at least some cases, observations demonstrate that the merger produces a final vortex larger than either of the two vortices that merged. Another relevant inference is the observation that small-scale eddies transport momentum into the jets \citep{Beebe80, Ingersoll81, Salyk06, DelGenio07}, which implies that the jet streams are maintained by quasi-organized turbulence at cloud level. By themselves, however, neither of these types of observation necessarily requires an inverse energy cascade.

An alternative approach for analyzing atmospheric energy transfer lies in the spectral analysis of the atmosphere's kinetic energy.  Because two-dimensional turbulence theory predicts specific slopes for the kinetic energy spectrum ($-5/3$ and $-3$ in the energy and enstrophy cascade regions, respectively), with a break in slope at the forcing scale, spectral analysis can help to answer the question of whether an inverse cascade exists and, of so, to determine the forcing scale.

To date, only a handful of studies have examined energy spectra of Jupiter's atmosphere. \citet{Mitchell82} performed the first direct measurement of the power spectrum of the turbulent kinetic energy derived from wind measurements. Using a one-dimensional fast Fourier transform (FFT), Mitchell measured the eddy zonal kinetic energy and determined that its average power spectrum slope is close to -1 but within the error bounds of the expected -5/3 value. However, his study utilized a regular grid of velocity values interpolated from scattered wind vector data. Furthermore, it was limited to only several jet stream latitudes and did not systematically examine the entire atmosphere. Later, \citet{Mitchell85} updated his earlier work to report average slopes of -1.3 and -3 in the two domains of the power spectrum (no uncertainty values were reported). Subsequent studies have performed a power spectrum analysis on Jovian cloud patterns with the intention of making inferences about turbulent energy transfer. \citet{Harrington96b} examined 1$^{\circ}$-wide zonal strips of cloud opacities in the near-infrared (5 $\mu$m) and determined their power spectra using Lomb normalization. For planetary wavenumbers greater than 25, they reported average power-law slopes of -3.14 $\pm$ 0.12 in westward jet streams and -2.71 $\pm$ 0.07 in eastward jet streams. For wavenumbers lower than 25, average power-law slopes were near -0.7 with negligible uncertainty for both westward and eastward jets. Recently, \citet{Barrado-Izagirre09} analyzed HST and \emph{Cassini} mosaics of Jupiter's atmosphere and calculated power spectra of zonal strips extracted from these mosaics using an FFT. In both the near-infrared and blue wavelengths, the average power spectrum slopes are -1.3 $\pm$ 0.4 and -2.5 $\pm$ 0.7 for the low and high wavenumber domains, respectively. Analysis of higher-altitude clouds observed using an ultraviolet filter returned slopes of -1.9 $\pm$ 0.4 and -0.7 $\pm$ 0.4, suggesting that either a different dynamical regime controls the atmosphere or that the relationship between cloud patterns and the embedded dynamics is different at this altitude. 

However, to the best of our knowledge, no study has performed a direct, extensive comparison using a common imaging data set of the cloud pattern power spectrum with the power spectrum of kinetic energy calculated using wind vectors derived from the same imaging data. (We note that \citet{Mitchell82} showed one power spectrum of a digitized albedo field at one jet stream and determined that its power spectrum reasonably fit a -5/3 power law.) Furthermore, we implement a spherical harmonic analysis to determine the energy spectrum of our 2D data sets, whereas previous studies typically analyzed 1D strips from their data with FFTs or other methods. We will demonstrate that analysis of 1D strips can affect the results and may not be directly comparable with turbulence theory. Using a modern data set from the \emph{Cassini} flyby of Jupiter in 2000 and our automated cloud feature tracking technique, we aim to constrain what relationship exists, if any, between the power spectrum of the turbulent kinetic energy and the power spectrum of cloud patterns in Jupiter's atmosphere.

\section{Data and Methodology}

\subsection{Description of the Data Set}

The Imaging Science Subsystem (ISS) \citep{Porco04} onboard \emph{Cassini} obtained an extensive multi-spectral data set of Jupiter's atmosphere in December 2000. We obtained the data set through the NASA Planetary Data System (PDS) Atmospheres Node. Throughout the particular observation sequence utilized here, the instrument alternated between observations of the northern and southern hemisphere of Jupiter, cycling through four spectral filters within each observation using its narrow angle camera (NAC). These filters were a near-infrared filter (CB2, 756 nm central bandpass), a visible blue filter (BL1, 455 nm), and two methane band filters (MT2, 727 nm; MT3, 889 nm). Before archival in the PDS, the raw images were calibrated with VICAR/CISSCAL 3.4 \citep{Porco04}, navigated using a planetary limb, and mapped onto a rectangular (simple cylindrical) projection. The CB2 images are mapped at a resolution of 0.05$^{\circ}$ pixel$^{-1}$ ($\approx$ 60 km pixel$^{-1}$), whereas the BL1, MT2, and MT3 images are mapped at a resolution of 0.1$^{\circ}$ pixel$^{-1}$ ($\approx$ 120 km pixel$^{-1}$). The image maps use the convention of System III positive west longitude and planetocentric latitude.

The extensive set of component image maps enabled us to create three global mosaics for each of the four spectral filters used in the observation sequence. Table \ref{Table: turb_mosaic_data} lists the ranges of \emph{Cassini} spacecraft clock times for the raw images from all four filters that we integrated into mosaics. Each mosaic is not a snapshot of the atmosphere at one instantaneous epoch, but an assemblage of images taken throughout one Jovian rotation. In addition, the temporal spacing between each mosaic of a certain filter is one Jovian rotation. To create each mosaic, we stitched together approximately 20 component images for each mosaic after applying a Minnaert correction to each component image to remove the limb-darkening and uneven solar illumination. To reduce the presence of seams in overlapping areas of the component images, we used a weighted average of the image brightness values at a particular overlap pixel. The value from each component image is weighted by their distances to the overlapping edges, which is the process employed by \citet{Peralta07} and \citet{Barrado-Izagirre09}.

\begin{table}
\begin{center}
\begin{tabular}{rrrr}
\hline
Filter & Mosaic & Spacecraft Clock Time (sec) & Longitude (System III) at left edge\\
\hline \hline

CB2 & 1 & 1355245203--1355279277 & 319.45 \\
CB2 & 2 & 1355279277--1355313351 & 319.45 \\
CB2 & 3 & 1355313351--1355351384 & 319.45 \\
BL1 & 1 & 1355245166--1355283026 & 356.8\\
BL1 & 2 & 1355282626--1355317100 & 356.8\\
BL1 & 3 & 1355316700--1355354960 & 356.8\\
MT2 & 1 & 1355233478--1355264170 & 170.4\\
MT2 & 2 & 1355263766--1355298244 & 170.4\\
MT2 & 3 & 1355297840--1355336104 & 170.4\\
MT3 & 1 & 1355233511--1355264203 & 170.4\\
MT3 & 2 & 1355263799--1355298277 & 170.4\\
MT3 & 3 & 1355297873--1355336137 & 170.4\\

\hline
\end{tabular}

\caption[Ranges of \emph{Cassini} spacecraft clock times for the component images composing our image mosaics] {\label{Table: turb_mosaic_data} Ranges of \emph{Cassini} spacecraft clock times for the component images composing our image mosaics.}

\end{center}
\end{table}

Independent of the global mosaics, we created image pairs for cloud tracking from the set of CB2 component images. CB2 images are well-suited for automated feature tracking because they possess high contrast. The frequency of observations produced redundant longitudinal coverage in consecutive observations that enabled us to create image pairs in these areas of overlap, and as a result, measure the advection of cloud features. We paired each component CB2 image with the next immediate one in the sequence, as this results in the most longitudinal overlap. The typical separation time between images in a pair is 63m 6s (3,786 s), though some pairs have shorter or longer separation times by $\sim$100 s. We compensated for differences in illumination within the image pairs using a Minnaert correction. For purposes of cloud tracking, we also applied a uniform high-pass filter to each image in the image pair to enhance the contrast of the cloud albedo pattern. (We did not perform such filtering on the image mosaics used to obtain power spectra of cloud brightness, however.) 

\subsection{Automated Feature Tracking}

We measured wind vectors for each image pair with our automated cloud feature tracker \citep{Choi07}. We used a square, 2$^{\circ}$ correlation box for the initial measurement. Our software then refines this measurement with a square, 0.5$^{\circ}$ correlation box. The cloud feature tracker returned robust results for most of the area in each image pair, resulting in component wind vector maps at a resolution of 0.05$^{\circ}$ pixel$^{-1}$, matching the resolution of the CB2 images. The set of approximately 70 image pairs provided sufficient longitudinal coverage that enabled us to create three independent full-longitudinal wind vector mosaics. The temporal spacing between these mosaics is one Jovian rotation. We employ the same weighted average scheme described earlier for stitching image albedo maps when connecting component wind vector maps to each other. We omitted vectors poleward of 50$^{\circ}$ latitude, as the quality of the wind vector results substantially decreased. We believe this is attributable to insufficient contrast and resolution at high latitudes because of observation geometry and image reprojection. 

One issue regarding our technique is that an individual wind vector map created by our software does not fully encompass the geographic area of an input component image pair, as wind measurements typically within a degree of the image edges are not possible. This is inconsequential for both the eastern and western edges of the individual wind vector component maps because these maps are stitched together across longitudes when creating mosaics. (In other words, individual wind vector component maps are spaced in longitude by less than their width in longitude, so that there is substantial longitude overlap between maps.)  It is also unimportant at high latitudes because the wind vector mosaics are cropped at 50$^{\circ}$ latitude. However, the remaining edge in the component maps is the equator; therefore, the component wind vector maps extend to only $\sim$1$^{\circ}$ latitude. When stitching the component maps together at the equator to create the global wind vector mosaics, we complete this equatorial gap by interpolating with weighted nearest-neighbor averaging.  The result is three independent wind vector maps extending from $50^{\circ}$S to $50^{\circ}$N and covering the full $360^{\circ}$ of longitude. 

The combination of imperfect navigational pointing knowledge of the images, the sizes of the correlation windows used for feature tracking, and the local wind shear all contribute to the uncertainty for each wind vector \citep{Choi07}. We estimate that the maximum uncertainty for an individual wind vector is $\sim$15 m s$^{-1}$, and that typical uncertainties are $\sim$5--10 m s$^{-1}$. In addition, our cloud tracking software returns obviously spurious results in areas of poor image resolution or imaging artifacts. We remove spurious results by employing a combination of bad pixel removal software (utilizing similar methodology as removing cosmic ray hits) and manual identification of bad vectors. From our experience, no objective method to identify spurious results has achieved satisfactory performance to date in recognizing every bad pixel. At most, $\sim$5\% of the measurements in a component wind vector map are spurious; typically, only 1--2\% of the vectors are removed. We fill in the locations flagged as spurious with nearest-neighbor averaging by defining the value of the wind vector at the location of a bad pixel to be the average of the vectors in its immediately surrounding pixels. 

\section{Cloud Albedo and Wind Vector Maps}

\begin{figure}[htbp]
  \centering
  \includegraphics[keepaspectratio=true, totalheight=7in]{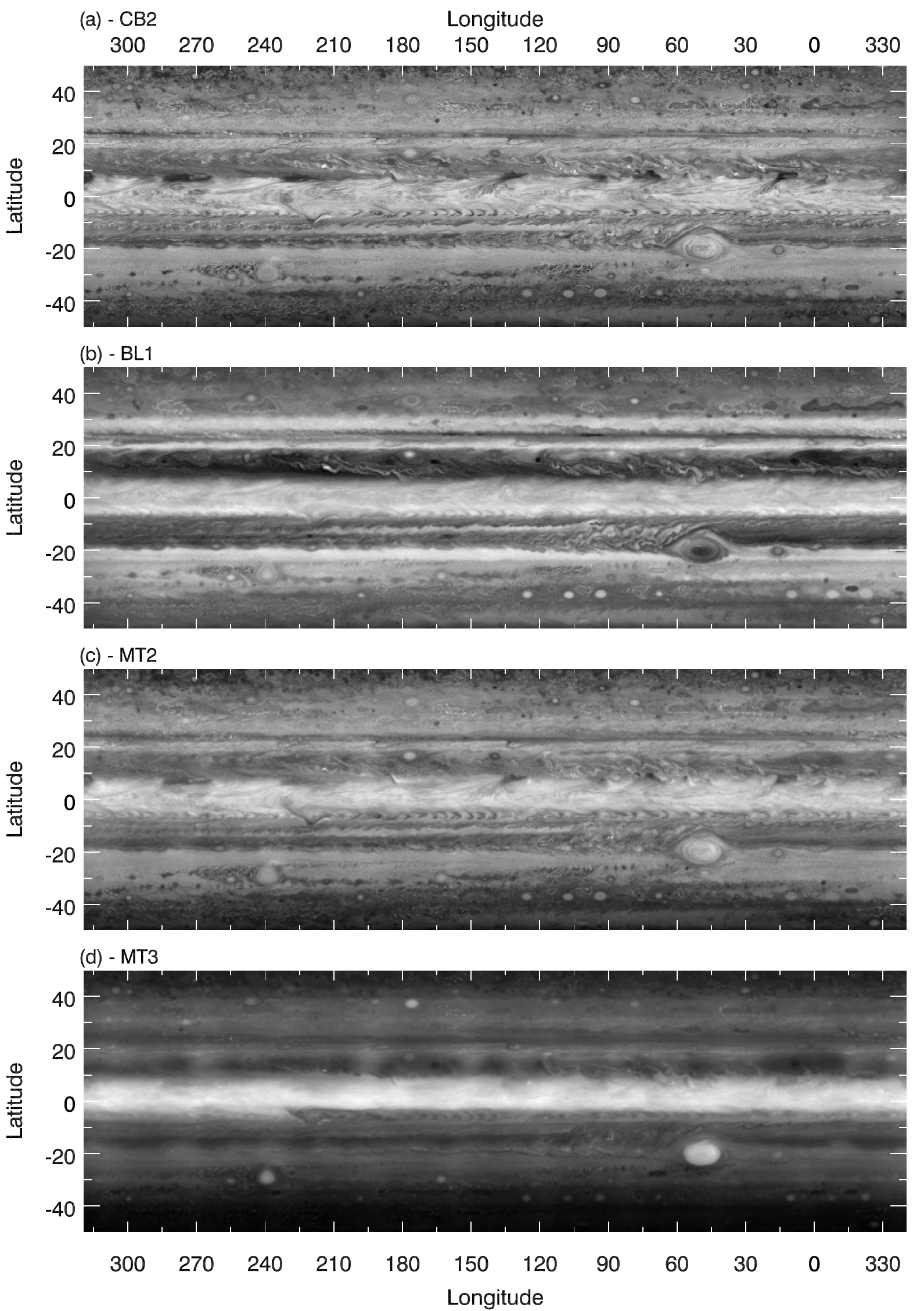}
  \caption[Jupiter's atmosphere as imaged by various filters on \emph{Cassini}, December 2000]{
    \label{Figure: union_all_imgs}
    Full-longitudinal mosaics of Jupiter's atmosphere and cloud structure composed with \emph{Cassini} ISS images taken over one Jovian rotation in December 2000. The resolution of this mosaic is 0.05$^{\circ}$ pixel$^{-1}$ ($\approx$ 60 km pixel$^{-1}$) for the CB2 mosaic (a) and 0.1$^{\circ}$ pixel$^{-1}$ ($\approx$ 120 km pixel$^{-1}$) for the remaining three mosaics (b-d).
    }
\end{figure}

Figure \ref{Figure: union_all_imgs} displays a mosaic observed using each of the spectral filters examined in this work. Observations using the CB2 filter are sensitive to deep layers ($>$ 10 bars) \citep{Banfield98, Porco03}. Many of Jupiter's well-known atmospheric structures are clearly evident in the CB2 (near-infrared) mosaic, including cloud bands latitudinally alternating in albedo, a diverse array of vortices, turbulent eddies, and dark equatorial hot spots. The structures seen in the CB2 mosaics are thought to reside within the ammonia cloud deck whose base is at an altitude around 700--1100 mbar \citep{Irwin98, Matcheva05, Sromovsky10}. BL1 (visible blue) images are also sensitive to deep layers but are affected by inherent color variations in the clouds and hazes as well as Rayleigh scattering \citep{Vasavada06}. As a result, the BL1 mosaic presents some striking differences with the near-infrared mosaic. The inherent reddish color of the equatorial belts (both centered at approximately 15$^{\circ}$ latitude) in each hemisphere as well as the interior of the Great Red Spot leads to a conspicuous darkening of these features in the BL1 mosaics. Also note that the lighter colored zones appear brighter in the BL1 mosaics, resulting in an increased dynamic brightness range within the blue mosaic. Furthermore, the dark areas associated with 5 $\mu$m hot spots are relatively indistinguishable in the BL1 mosaic compared to the CB2 mosaic. 

The methane bands can observe atmospheric haze layers located in the upper troposphere and stratosphere \citep{West86}. The MT3 methane filter is ideal for measuring stratospheric aerosols as it is sensitive to down to the upper troposphere ($\sim$300 mbar), whereas the MT2 methane filter can detect a denser upper tropospheric haze layer and cloud tops as these observations are sensitive to pressures up to 4 bars \citep{Banfield98}. The MT2 methane band mosaic looks very similar to the CB2 mosaic, but with subdued contrasts relative to both the BL1 and CB2 mosaics. Finally, the MT3 methane band mosaic is the most unique of the set, as many detailed cloud patterns seen in other mosaics are not visible in the MT3 mosaic. This is consistent with \citet{Banfield98}, who determined that the higher altitude haze layers had little to no lateral variability at length scales less than those of the planetary jet streams.  

\begin{figure}[htbp]
  \centering
  \includegraphics[keepaspectratio=true, totalheight=7in]{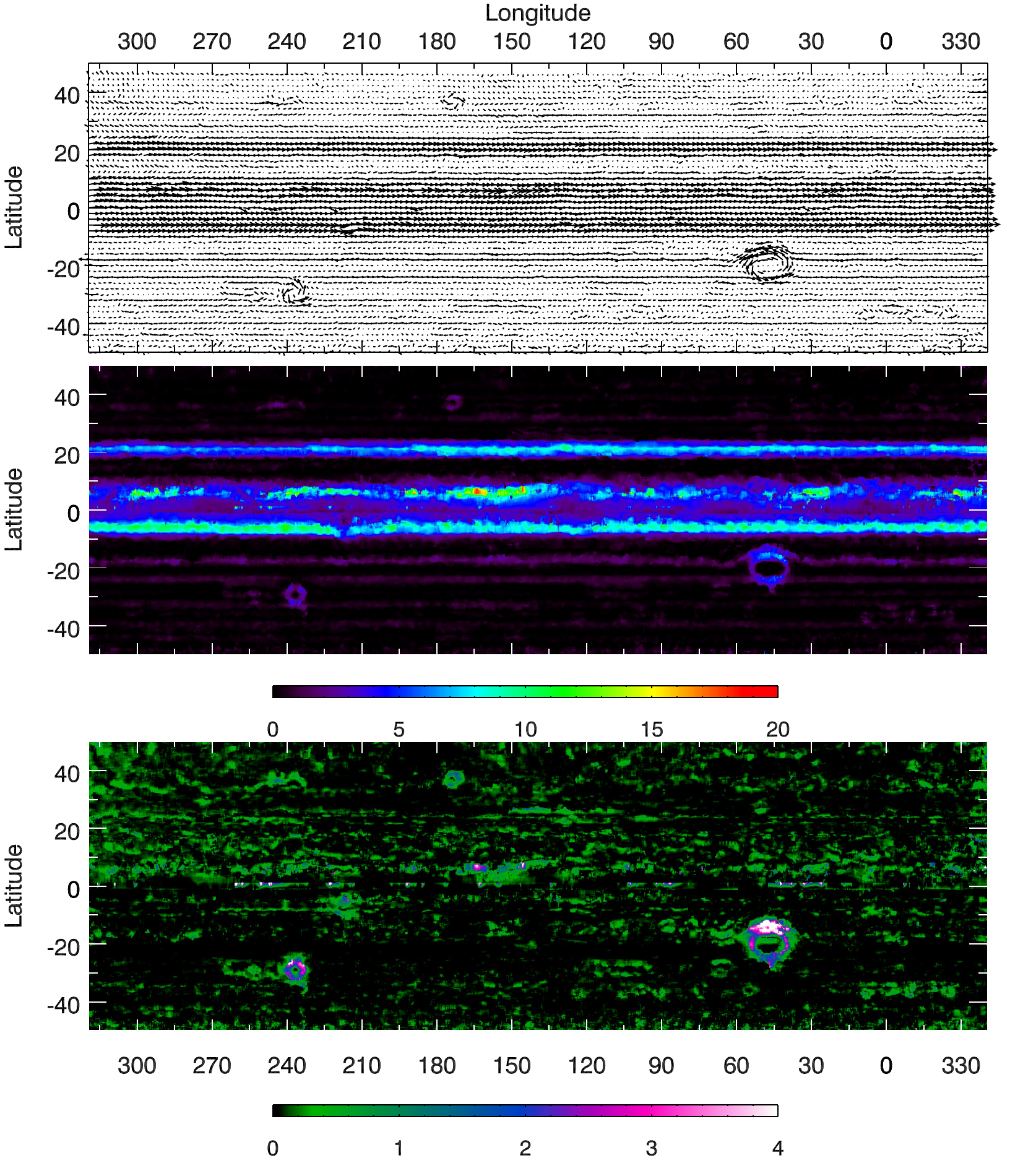}
  \caption[Wind vectors, total kinetic energy, and eddy kinetic energy of Jupiter's atmosphere, December 2000]{
    \label{Figure: union01}
    \emph{Top}: Wind vector mosaic \#1 determined by our automated feature tracker using image pairs generated from component images comprising the mosaics. A small percentage ($<$ 0.1\%) of vectors are shown here for clarity.  \emph{Middle}: Color map \#1 of the total kinetic energy per unit mass in Jupiter's atmosphere. \emph{Bottom}: Color map \#1 of the eddy kinetic energy per unit mass in Jupiter's atmosphere. The units of both scale bars are 10$^3$ m$^2$ s$^{-2}$.
    }
\end{figure}

Figure \ref{Figure: union01} shows one of our wind vector maps, along with corresponding maps of the total and eddy kinetic energy. (Our definition of eddy kinetic energy is in the following section.) The wind vector map reveals that the flow in the atmosphere is strongly banded and dominated by the jets, with only the circulation of the largest vortices readily apparent in the wind vector map in Figure \ref{Figure: union01}. The total kinetic energy map reveals that the broad equatorial jet and the northern tropical jet stream just north of 20$^{\circ}$N contain most of the kinetic energy in the troposphere. Some of Jupiter's other jet streams are also visible, along with other prominent vortices in the atmosphere such as the Great Red Spot and Oval BA. Smaller areas of circulation, eddy transport, and other latitudinal mixing that is strongly implied to occur between jets are not readily visible given the constraints of the wind vector and kinetic energy maps.
 
Though the Great Red Spot and Oval BA appear relatively insignificant compared to Jupiter's equatorial and north tropical jet streams in the total kinetic energy map, these two vortices have significant eddy kinetic energy. Other vortices and closed circulations are less prominent but visible in the mid to high latitudes. Longitudinal perturbations in kinetic energy are not exclusive to vortices, however. Both kinetic energy maps in Figure \ref{Figure: union01} reveal that the equatorial jet, and to a lesser degree, the 23$^{\circ}$N tropical jet, exhibit prominent longitudinal variance in kinetic energy. \citet{Garcia-Melendo10} report similar variations in their examination of the dynamics within Jupiter's equatorial atmosphere. We are confident that these perturbations within the jet streams are real because they are consistently present at approximately the same longitudes in our three independent velocity maps. In addition, the level of variance is greater than our estimated uncertainty in the velocity maps: the degree of variation in zonal velocity within these jet streams is typically 20--30 m s$^{-1}$ and in some cases, much higher.

\begin{figure}[htbp]
  \centering
  \includegraphics[keepaspectratio=true, width=5.9in]{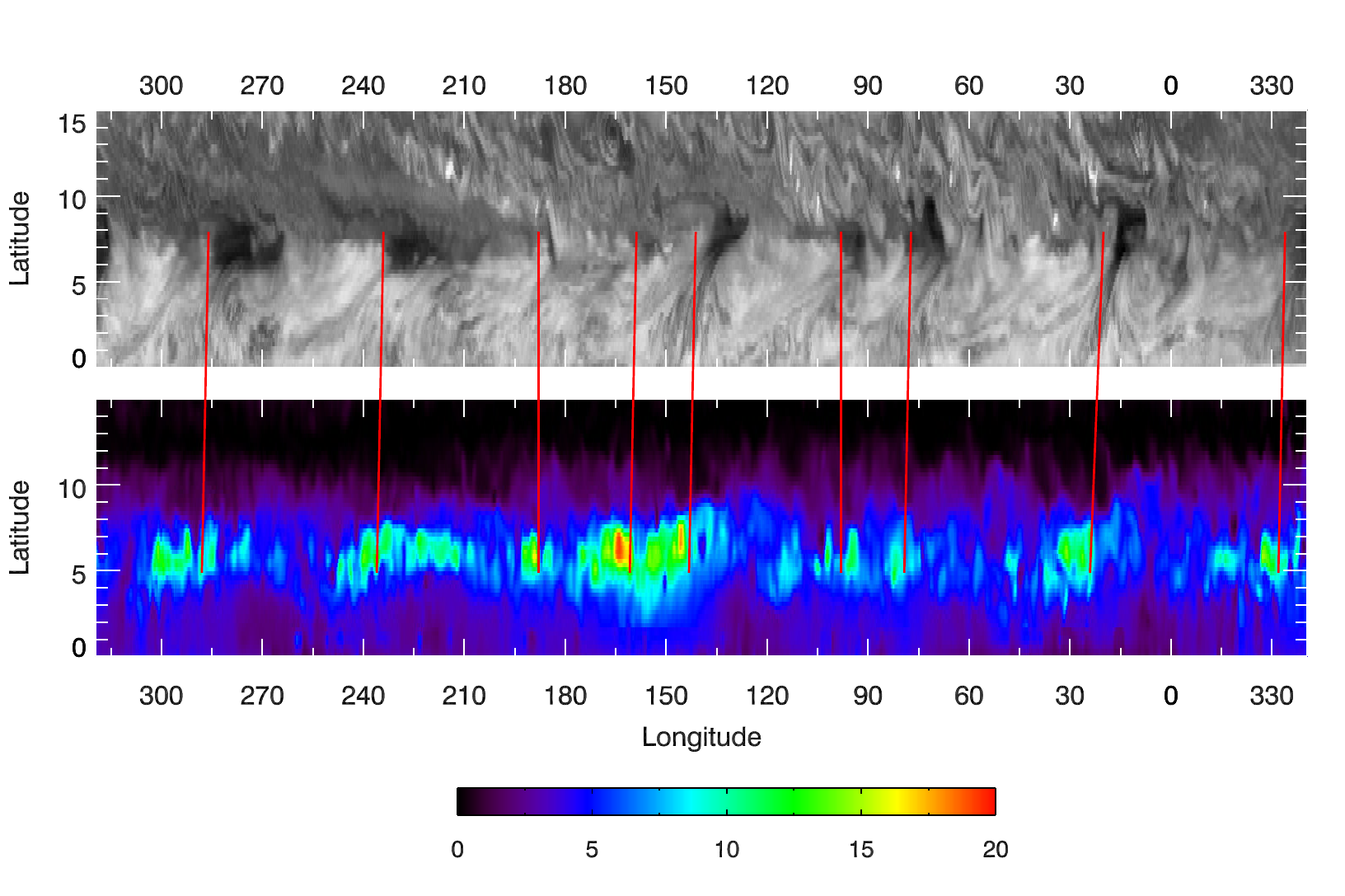}
  \caption[Comparison of Jupiter's North E	quatorial Jet Stream in near-infrared and Total Kinetic Energy]{
    \label{Figure: hotspot_closeup}
    Comparison of Jupiter's North Equatorial Belt in near-infrared (\emph{top}) and total kinetic energy (\emph{bottom}). White lines mark the western edge of visible hotspots with the eastern edge of kinetic energy perturbations. The units of the scale bar are 10$^3$ m$^2$ s$^{-2}$.
    }
\end{figure}

This variance within the jets is consistent with the presence of large-scale atmospheric waves that may play a role in controlling the dynamics of the equatorial jet stream. Current models for the generation and maintenance of Jupiter's 5 $\mu$m hotspots invoke the presence of an equatorially trapped Rossby wave \citep{Allison90, Ortiz98, Showman00, Friedson05}. The strongest signature for these waves appears at the 7.5$^{\circ}$N jet stream, which corresponds to the latitude of the equatorial hotspots \citep{Orton98}. As shown in Figure \ref{Figure: hotspot_closeup}, we find that positive perturbations in kinetic energy lie immediately to the west of the hotspots, which appear as periodic dark patches between 5$^{\circ}$N and 10$^{\circ}$N. This broadly matches the observations made by \citet{Vasavada98}, who described strong flow entering a hotspot at its southwestern periphery, though the flow exiting the hotspot remained unconstrained by their work. 

\begin{figure}[htbp]
  \centering
  \includegraphics[keepaspectratio=true, width=5.9in]{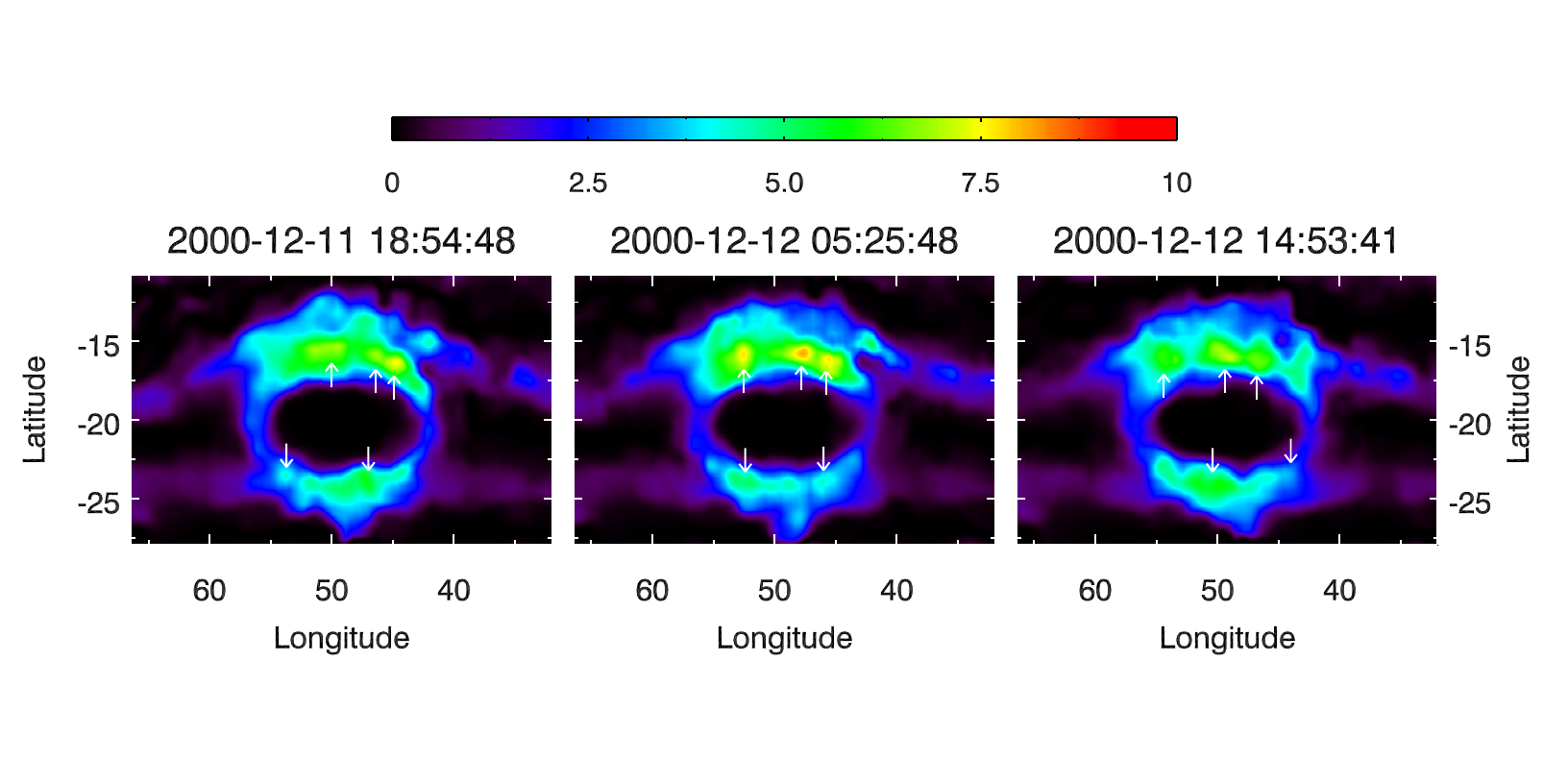}
  \caption[Closeup time series of the kinetic energy per unit mass of the Great Red Spot]{
    \label{Figure: union_grs_eddy_closeup}
    Time series of the kinetic energy per unit mass, zoomed in on Jupiter's Great Red Spot. The arrows denote local regions of anomalous kinetic energy that appear to strengthen or dissipate during the observation sequence. Listed times denote the observation time in UTC of the first image of an image pair used for wind measurement. The units of the scale bar are 10$^3$ m$^2$ s$^{-2}$.
    }
\end{figure}

We also report prominent variance in the kinetic energy within the main circulation collar of the GRS. Figure \ref{Figure: union_grs_eddy_closeup} is a detailed map displaying these eddies embedded within the high-velocity flow collar of the GRS. The magnitude of the velocity anomaly for these eddies is 10--30 m s$^{-1}$ above their immediate surroundings (the northern portion of the GRS flow collar). These eddies appear to fluctuate in strength and may be advecting with the winds in the flow collar. In our data set, areas of stronger flow are present at the northern half of the vortex during the \emph{Cassini} flyby, matching previous work by \citet{Asay-Davis09}, who used the same \emph{Cassini} data set as this current work. In contrast, nearly seven months prior to these \emph{Cassini} observations, \emph{Galileo} observed the GRS at high resolution during its G28 orbit in May 2000 and found stronger flow at the \emph{southern} half of the vortex \citep{Choi07}. The contrast in the location of the region of stronger flow in the GRS implies that eddies or waves influence the dynamical meteorology of large Jovian vortices, and that velocity maps or profiles built from observational snapshots may not be representative of a time-mean state of these meteorological features.

\section{Power Spectra}

\subsection{Spherical Harmonic Analysis}

We computed the power spectra of the total kinetic energy per unit mass and the illumination-corrected cloud brightness patterns observed at different wavelengths using spherical harmonic analysis. We also computed the power spectra of the eddy kinetic energy per unit mass, which we define as $\mathit{KE}_{\mathrm{eddy}} = \frac{1}{2}\left(u'^2 + v'^2\right)$, where $u'$ and $v'$ are the eddy components of zonal and meridional velocity, respectively. The eddy component of the zonal velocity is the observed velocity $u$ without the zonal mean contribution ($u' = u - \overline{u}$). Instead of using a previously calculated zonal-wind profile, we define the zonal mean of the zonal-wind ($\overline{u}$) from the derived velocities: $\overline{u}$ for each discrete, full-longitudinal row of pixels is simply the mean value of $u$ in the row. We use a similar process to determine $v'$ and $\overline{v}$. (Although it is expected to be small, we do not assume that $\overline{v}$ is zero.)

To calculate the power spectrum, we first perform a spherical harmonic analysis of the data using NCL (NCAR Command Language, www.ncl.ucar.edu). This yields a map of spectral power as a function of both total wavenumber $n$ (the lower coefficient of the spherical harmonic) and zonal wavenumber $m$. We calculate the total power within each discrete total wavenumber by integrating the power across all zonal wavenumbers $m$. Once we calculate the one-dimensional power spectrum in this way, we determine the best power-law relationship through linear least-squares fitting of the spectrum as a function of total wavenumber. Our fitting algorithm calculates the best-fit slope of the spectrum, defined as the value of the exponent $k$ when fitting the energy spectrum to a power law $E(n) = E_on^{-k}$ across total wavenumbers $n$. Our algorithm extracts a portion of a power spectrum bounded at two discrete wavenumber boundaries that are free parameters. We set these parameters at wavenumber 10 for the lower bound and wavenumber 1000 for the upper bound. Our algorithm also has two modes: a ``single slope'' mode for fitting one slope to the entire extracted power spectrum, and a ``composite slope'' mode for fitting two independent slopes---one at low wavenumbers and one at high wavenumbers---to the extracted spectrum. In the composite slope mode, we do not arbitrarily assume a wavenumber at which the transition in slope occurs. Instead, we automatically determine the location of this transition by calculating the two independent best fit slopes for each possible transition point; the overall best fit is the one with the lowest $\chi^2$. The uncertainty in transition location corresponds to the range of transitions where the fitted slopes are within the uncertainties in each slope established by the overall best fit. The current analysis only examined spectral slopes; the spectral amplitude is also an important property for characterizing power spectra but is left for future studies. 

Spherical harmonic analysis using NCL requires a complete global data set as input to return valid results. All of our image mosaics are fully global, though some small portions of each mosaic near the poles are missing because of a lack of observational coverage. This has a negligible effect on our results. Our wind vector and kinetic energy maps, however, omit results higher than 50$^{\circ}$ latitude. We fill in the missing wind data at high latitudes with the \citet{Porco03} zonal-wind profile, and assume that the winds at these latitudes are constant with longitude. This choice neglects the small-scale longitudinal wind structure that surely exists poleward of $50^{\circ}$ latitude and hence underestimates the globally integrated spectral power at the largest wavenumbers. To determine whether this affects the global spectrum, we also experimented with inserting plausible profiles of longitudinal variability poleward of $50^{\circ}$ latitude, which show that such variability has only a very minor effect on the spectrum.  These sensitivity studies are further discussed in Section~\ref{hld}.

\begin{figure}[htbp]
  \centering
  \includegraphics[keepaspectratio=true, totalheight=6.5in]{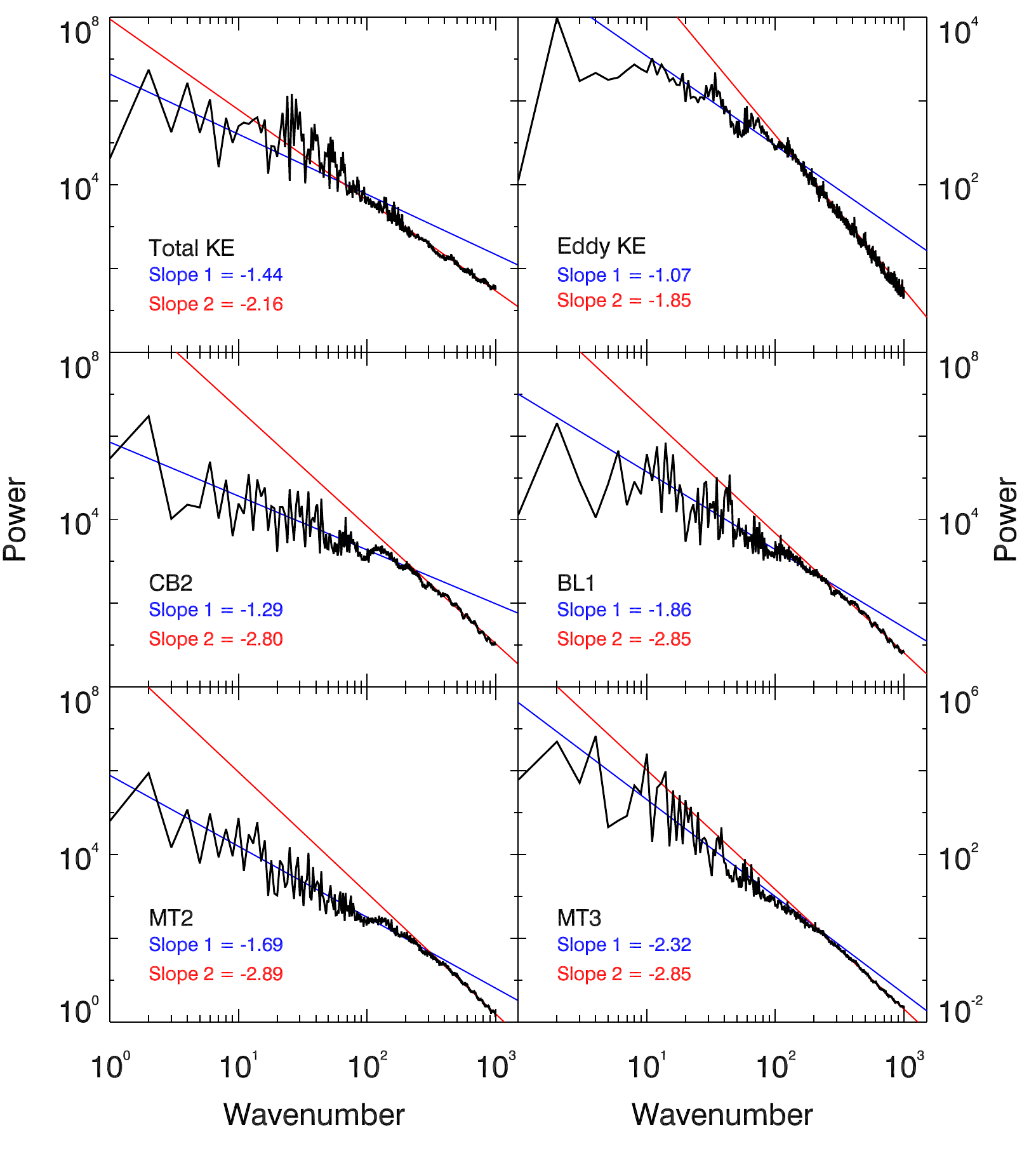}
  \caption[Power Spectra of Selected Data Sets]{
    \label{Figure: sph_powspec}
    Power spectra for selected data sets from our study. The solid lines are best-fit slopes for each power spectrum when assuming a composite fit. The algorithm searches for the transition in the slope between wavenumbers 10 and 1000. 
    }
\end{figure}

Figure \ref{Figure: sph_powspec} plots total wavenumber $n$ versus spectral power for representative power spectra calculated from the data sets in our study. In addition, Table \ref{Table: sph_psd_slopefit} lists the best-fit slopes for all of our calculated power spectra. In all cases, fitting with a composite of two slopes is the quantitatively preferred fit, with a transition in slope between wavenumbers 50 and 350. We did not calculate power spectra beyond total wavenumber 1000, as spectral power beyond this is negligible to the overall spectrum. Aliasing and other spectral effects are not an issue as the Nyquist frequency of the input data corresponds to a planetary wavenumber of $\sim$3,500 for the CB2 and kinetic energy maps at 0.05$^{\circ}$ pixel$^{-1}$, and $\sim$1,750 for the BL1, MT2, and MT3 mosaics at 0.1$^{\circ}$ pixel$^{-1}$.

At very low wavenumbers (n $<$ 10), the kinetic energy and image mosaic spectra do not exhibit power law behavior but rather spike up and down, with spectral power typically 1--2 orders of magnitude greater at even wavenumbers than odd wavenumbers. This behavior results from the fact that, at low wavenumbers, the zonal jet streams dominate the spectra, and because the jets are approximately symmetric across the equator, the power at even wavenumbers greatly exceeds that at odd wavenumbers. (If the jets were precisely symmetric about the equator, the power at odd wavenumbers would be zero; the small but nonzero power at odd wavenumbers less than 10 in the real spectrum results from the modest degree of asymmetry about the equator.) The deviation of the spectrum from a power law at small wavenumbers therefore makes sense. At large wavenumbers, in contrast, the spectrum averages over the individual signatures of hundreds or thousands of small eddies and, because of this averaging, can produce a smooth power law. Accordingly, when fitting a power-law to our spectra, we only consider the portion of the spectra at wavenumbers greater than 10.

The power spectra of the image mosaics resemble the theoretical power spectrum predicted by forced two-dimensional turbulence theory: typical slopes in the low wavenumber domain of the image mosaic power spectrum (Slope 1) are near -5/3, and slopes in the high wavenumber domain (Slope 2) approach -3. The lone exception is the power spectrum of the strong methane band (MT3) mosaic, which exhibits a steeper slope (near -2.3) at low wavenumbers. This is expected given the observed paucity of small-scale structure in the clouds and haze in this particular mosaic, as the methane band filters sample higher altitudes that are likely subject to other dynamical regimes and chemical processes. The location of the transition in slopes for the power spectra across all image mosaics typically range from 200--300 in total wavenumber, corresponding to a length of $\sim$1,500--2,000 km. The remarkably similar slopes for the image mosaics in the high wavenumber region across all four observed spectral bands is somewhat surprising. This suggests that the level of contrast variations at these local length scales are similar for all four spectral filters examined in this study. Our results indicate, however, that the cloud contrast variations at longer length scales (lower wavenumbers) do affect the power spectra.

The kinetic energy spectra also exhibit important distinctions from the image mosaic spectra. These differences are robust; as shown in Table 2, the spectral slopes are relatively consistent across the three independent data sets for each type of image mosaic and kinetic energy data set in our analysis. Though the low wavenumber region of the total kinetic energy spectra exhibit slopes near -5/3, similar to the image mosaics, the eddy kinetic energy spectra typically have flatter slopes, near -1, at low wavenumbers. Furthermore, in the high wavenumber regions of both eddy and total kinetic energy spectra, best-fit slopes are typically shallower than those of the cloud brightness, with typical slopes near -2. The wavenumber of the slope transition is markedly smaller than that in the image mosaic spectra. For the total kinetic energy spectra, this transition is near wavenumber 70 (wavelength $\sim$6,000 km), while the transition is near wavenumber 140 for the eddy kinetic energy spectra (length $\sim$3,000 km). The location of the transition in slope is typically interpreted as the length scale at which energy is injected into the atmosphere. However, this interpretation seems difficult to reconcile with the fact that the transition wavenumbers in kinetic energy differ significantly from those of the images from which the kinetic energy is derived. The precise choice of slope fits to our spectra affect the value of the crossover wavenumber, and so for a given spectrum, the transition wavenumber exhibits some uncertainty, probably at the factor-of-two level.  Thus, the distinct transition wavenumbers obtained from kinetic energy and image brightness spectra may actually be consistent within error bars.  Alternatively, it is possible that dynamical instabilities (e.g. moist convection or baroclinic instabilities) exhibit differing dominant length scales for their kinetic energy and clouds; in this case, global kinetic energy and cloud image spectra would indeed exhibit differing transition wavenumbers.

\begin{table}
\begin{center}
\begin{tabular}{|l|c|c|c|}

\hline
 & Slope 1 & Slope 2 & Transition \\
\hline

KE 1 & -1.44 $\pm$ 0.27 & -2.16 $\pm$ 0.01 & 67$^{+6}_{-2}$\\
KE 2 & -1.36 $\pm$ 0.27 & -2.17 $\pm$ 0.01 & 67$^{+6}_{-1}$\\
KE 3 & -1.35 $\pm$ 0.26 & -1.97 $\pm$ 0.01 & 67$^{+6}_{-2}$\\ 
 & & & \\
Eddy KE 1 & -1.07 $\pm$ 0.03 & -1.85 $\pm$ 0.01 & 142$^{+15}_{-10}$\\
Eddy KE 2 & -1.06 $\pm$ 0.03 & -1.91 $\pm$ 0.01 & 142$^{+14}_{-8}$\\
Eddy KE 3 & -0.86 $\pm$ 0.03 & -1.74 $\pm$ 0.01 & 138$^{+11}_{-6}$\\ 
 & & & \\
CB2 1 & -1.29 $\pm$ 0.05 & -2.80 $\pm$ 0.01 & 227$^{+37}_{-25}$\\
CB2 2 & -1.28 $\pm$ 0.05 & -2.73 $\pm$ 0.01 & 229$^{+13}_{-16}$\\
CB2 3 & -1.30 $\pm$ 0.05 & -2.74 $\pm$ 0.01 & 232$^{+11}_{-12}$\\ 
 & & & \\
BL1 1 & -1.86 $\pm$ 0.05 & -2.86 $\pm$ 0.01 & 244$^{+19}_{-18}$\\
BL1 2 & -1.87 $\pm$ 0.05 & -2.79 $\pm$ 0.01 & 257$^{+16}_{-27}$\\
BL1 3 & -1.86 $\pm$ 0.05 & -2.77 $\pm$ 0.01 & 243$^{+29}_{-20}$\\ 
 & & & \\
MT2 1 & -1.69 $\pm$ 0.03 & -2.89 $\pm$ 0.01 & 303$^{+10}_{-6}$ \\
MT2 2 & -1.69 $\pm$ 0.03 & -2.91 $\pm$ 0.01 & 307$^{+10}_{-6}$ \\
MT2 3 & -1.70 $\pm$ 0.03 & -3.00 $\pm$ 0.01 & 321$^{+5}_{-5}$ \\ 
 & & & \\
MT3 1 & -2.32 $\pm$ 0.05 & -2.86 $\pm$ 0.01 & 202$^{+9}_{-3}$ \\
MT3 2 & -2.31 $\pm$ 0.04 & -2.87 $\pm$ 0.01 & 227$^{+26}_{-4}$ \\
MT3 3 & -2.29 $\pm$ 0.03 & -3.22 $\pm$ 0.01 & 313$^{+12}_{-10}$ \\ 

\hline

\end{tabular}

\caption[Power Spectrum slopes for the Image Mosaics and Kinetic Energy Maps]{
	\label{Table: sph_psd_slopefit}
	\label{lasttable}
	Best-fit slope values for the power spectra of the various data sets in our study. We report the values of the two slopes found by the composite fit, the 1-$\sigma$ uncertainty value for each slope fit, and the planetary wavenumber where the transition in slope occurs. The slope of the power spectrum is defined as the value of the exponent $k$ when fitting the spectrum to a power law $E(n) = E_on^{-k}$ across wavenumber $n$.}
\end{center}
\end{table}

\begin{figure}[htbp]
  \centering
  \includegraphics[keepaspectratio=true, width=5.5in]{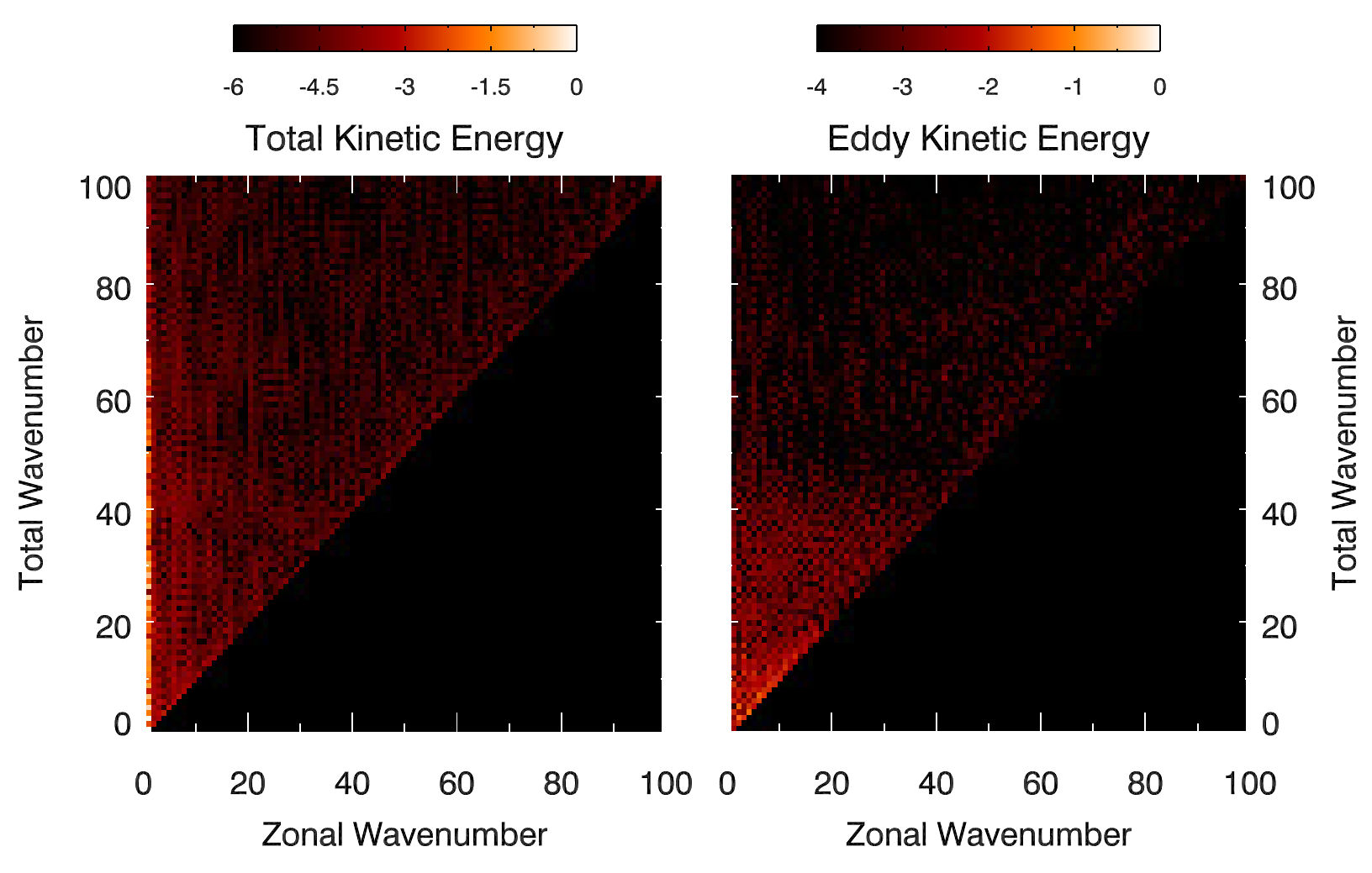}
  \caption[2D Power Spectrum]{
    \label{Figure: 2D_powspecmap}
    Power spectrum for total and eddy kinetic energy, Map \#1. We normalize each map with the maximum value in each spectrum and plot the base-10 logarithm of the normalized value. For clarity, we display each spectrum with independent scale bars. 
    }
\end{figure}

Figure \ref{Figure: 2D_powspecmap} maps spectral power as a function of zonal and total wavenumber. The zonal symmetry of Jupiter's jet streams produce anisotropy, strongly confining the spectral power of total kinetic energy to low zonal wavenumbers. The energy spectrum of eddy kinetic energy removes the dominant contribution of the zonal jet streams, and reveals enhanced power between total wavenumbers 1 and 40. However, our 2D energy spectra lack any such exclusion region; rather, to zeroth order, spectral power of eddy kinetic energy is nearly independent of zonal wavenumber, as would be expected if the eddy kinetic energy is isotropic. Qualitatively, this area of enhanced power resembles spectra of 2D turbulence simulations performed on a sphere \citep{Nozawa97, Huang98}. Such spectra also feature intensified power at low wavenumber. However, an important difference emerges between their work and our result: they predicted, based on simple scaling arguments and their turbulence simulations, that energy would be excluded from a portion of the spectrum at low wavenumber where Rossby waves rather than turbulence dominate (see Figs. 1--2 in \citet{Huang98}). This may provide evidence that interaction of an inverse energy cascade with Rossby waves is not occurring within Jupiter's atmosphere, or it could imply that nonlinear wave-wave interactions drive energy into that region. However, the non-conformity of our results with theory could also imply that vertical motions are important for the real interaction of turbulent eddies in Jupiter's atmosphere, and that simple 2D theory is not applicable. 

\subsection{Comparison with Previous Work}

Previous studies have calculated power spectra derived from data averaged into one-dimensional functions of latitude or longitude. Typically, this approach utilizes an FFT or a Legendre transform and produces spectra that differ substantially from spectra calculated from a full two-dimensional data set. For example, \citet{Galperin01} and \citet{Barrado-Izagirre09} calculated power spectra using latitudinal profiles of Jupiter's kinetic energy and image mosaics, respectively. \citet{Galperin01} used a zonally averaged, zonal-wind profile, smoothed in latitude, as a basis for a kinetic energy profile and determined its spectrum using spherical harmonic analysis. \citet{Barrado-Izagirre09} collapsed the observed brightness of their HST and \emph{Cassini} image mosaics into near-global one-dimensional latitudinal profiles by averaging across all available longitudes, and calculated spectra with a 1D latitudinal FFT. Both studies reported that their meridional spectra followed a power-law relationship with a slope ranging from -4 to -5 with no clear transition in slope. 

To determine the effect that a 2D data set has on resulting power spectra, we used zonal-mean zonal-wind profiles from \emph{Voyager} \citep{Limaye86}, \emph{HST} \citep{Garcia-Melendo01}, and \emph{Cassini} \citep{Porco03} as the basis for zonally symmetric maps of kinetic energy, which we then analyzed with our technique for calculating a 2D power spectrum. For comparison, \citet{Galperin01} used a zonal-mean zonal-wind profile from \emph{Voyager}, which was smoothed in approximately 0.25$^{\circ}$ bins. 

\begin{figure}[htbp]
  \centering
  \includegraphics[keepaspectratio=true, totalheight=5.5in]{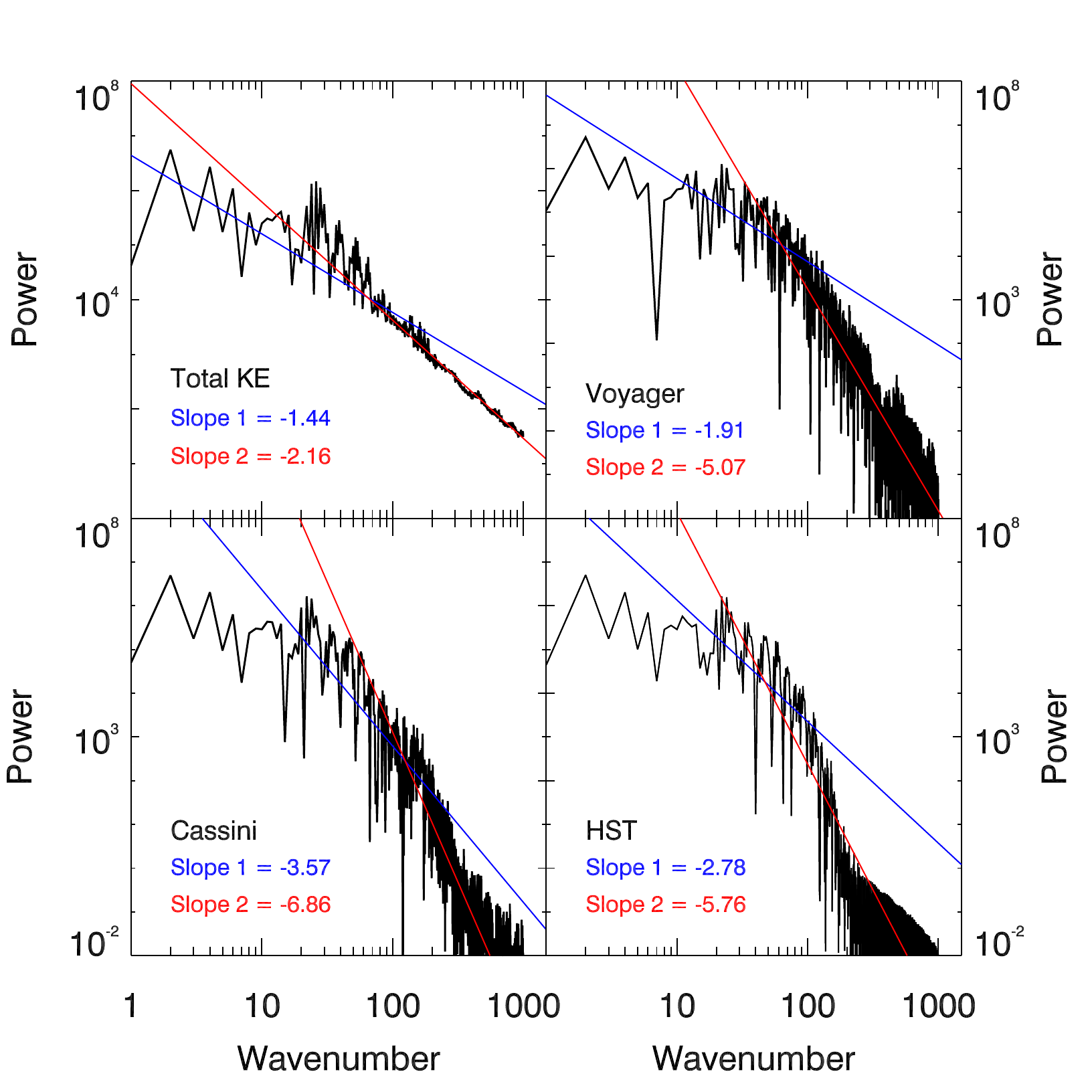}
  \caption[Power Spectra of Zonal Wind Profiles]{
    \label{Figure: sph_profile_spec}
    Power spectra of kinetic energy maps derived from zonal-wind profiles from \emph{Voyager}, \emph{Cassini}, and \emph{HST}, with the power spectrum of our kinetic energy map derived from full 2D velocity map cloud tracking data for comparison. The solid lines are representative best-fit slopes for an arbitrary domain in wavenumber when assuming a transition in slope exists. 
    }
\end{figure}

Our analysis demonstrates that power spectra calculated based on a zonal-mean zonal-wind profile, such as those examined by \citet{Galperin01}, differ significantly from spectra calculated from a two-dimensional map of wind vectors. Figure \ref{Figure: sph_profile_spec} displays these spectra, along with a power spectrum calculated from one of our 2D kinetic energy maps for comparison. The power spectrum we calculated from the \emph{Voyager} zonal-mean zonal winds is remarkably similar to the spectrum calculated by \citet{Galperin01}, as both feature a spectral slope near -5 at high wavenumbers. The detailed shape of spectra in figure \ref{Figure: sph_profile_spec} are all similar in the low wavenumber regions of the power spectrum (i.e. wavenumbers less than 10), as the dominant contribution to each spectrum here is made by the structure of the zonal jet streams, and this information is retained in both the 2D maps and 1D profiles. However, the spectra diverge at higher wavenumbers---all of the spectra calculated from 1D zonal-wind profiles exhibit significantly steeper slopes than those calculated from a 2D velocity map. 

Zonal-mean zonal-wind profiles used to calculate the 1D spectra do not account for eddies and turbulence present in the atmosphere. This implies that the spectral power associated with the longitudinal structure is neglected from the spectrum. Because small-scale structure typically dominates over the zonal-mean structure at high wavenumbers, this method significantly reduces the spectral power and steepens the spectral slopes in these regions of the spectrum. Therefore, a spectrum calculated from a 1D velocity profile is not representative of the total energy spectrum for wavenumbers $>$ 10.

Nevertheless, several studies have highlighted the useful role that spectra calculated from 1D wind profiles can play when characterizing anisotropic, turbulent flows. The $\beta$-effect, along with the interaction of turbulence and Rossby waves, generate signiÞcant anisotropy and can therefore alter the upscale transfer of energy in the zonal direction relative to the meridional direction \citep{Rhines75, Galperin06}. The production of zonal jets themselves is the ultimate manifestation of such anistropy. As emphasized by recent authors in the context of idealized 2D flows, the turbulence accompanying zonal jets necessarily also contains significant anistropy, at least over a range of wavenumbers (e.g, \citet{Huang01}; \citet{Galperin06, Galperin10}; \citet{Sukoriansky05, Sukoriansky07, Sukoriansky08}). For example, these authors demonstrate the existence of a flow regime, dubbed the zonostrophic regime, that can occur in 2D flows when (among other conditions) friction is sufficiently weak and there exists a sufficient scale separation between the peak energy-containing scale, the scale above which the turbulence becomes anisotropic, and the forcing scale; they showed that this regime can exhibit a steep -5 slope in the kinetic energy spectrum of the zonal modes (i.e., modes with zonal wavenumber m=0). At wavenumbers where these zonal modes are only a fraction of the total energy, this behavior would not be evident in a power spectrum of total kinetic energy calculated from a 2D velocity map. Still, as described by \citet{Galperin10}, one might expect the -5 behavior to dominate the \emph{total} (not just zonal) kinetic energy spectrum over some range of wavenumbers (e.g., see their Fig. 5). Our total kinetic energy spectra of Jupiter agrees with a wide range of slopes in the range of wavenumbers between 20 and 80, though we note that it is interesting that the upper envelope corresponds to a slope near -5. The short span of the data collection (only 3 consecutive Jovian days) is insufficient, however, for statistical validation of this result. 

\section{Sensitivity of Results to Noise, Smoothing, and High Latitude Data}

We investigated the consequences of several data processing techniques on the resulting power spectra. First, we investigated the effect of measurement noise on the spectra. Next, although we did not smooth our data at any point, previous studies have employed smoothing in their analysis. We demonstrate here the effects that smoothing has on the power spectrum. Finally, we examine our assumption of no longitudinal variability within the high latitude winds, and explore the sensitivity of the power spectrum to the choice of winds poleward of $50^{\circ}$ latitude where we lack wind data.

\subsection{Measurement Noise}

Measurement noise from our feature tracking technique is unavoidable. If measurement noise from our feature tracker is substantial, our calculated power spectrum would be perturbed at high wavenumbers so that best-fit slopes would be flatter (lower in magnitude) than best-fit slopes of the true spectrum. We examined the spectral contribution of measurement noise to eliminate the possibility that the noise adversely affected the results. Our wind vector map is the sum of the true wind and noise from our measurement technique. Similarly, the spectrum of our wind vector map is the sum of the spectrum of the true wind and the spectrum of the noise. By generating synthetic data with similar characteristics as the expected measurement noise and taking its spectra, we can determine whether or not the spectral contribution of the noise would affect the overall spectrum. 

We expect that the maximum uncertainty in any given wind measurement is between 10 and 20 m s$^{-1}$, and typical uncertainties are $\sim$5 m s$^{-1}$. We tested cases of random pixel-to-pixel noise at various amplitudes, some exceeding the expected levels of measurement noise by a substantial margin (e.g. rms values of noise approaching 30--40 m s$^{-1}$). However, we also expect that the wind vector map contains measurement noise correlated at a discrete length scale. This could be present in areas of cloud features with low contrast or in areas with high ambient wind shear, resulting in areas of vectors that possess similar uncertainties. We tested the effect of such noise by generating a random array of noise on a specified characteristic length scale and comparing the power spectra of these synthetic noise fields to that of our observational 2D wind field.

\begin{figure}[htbp]
  \centering
  \includegraphics[keepaspectratio=true, width=4in]{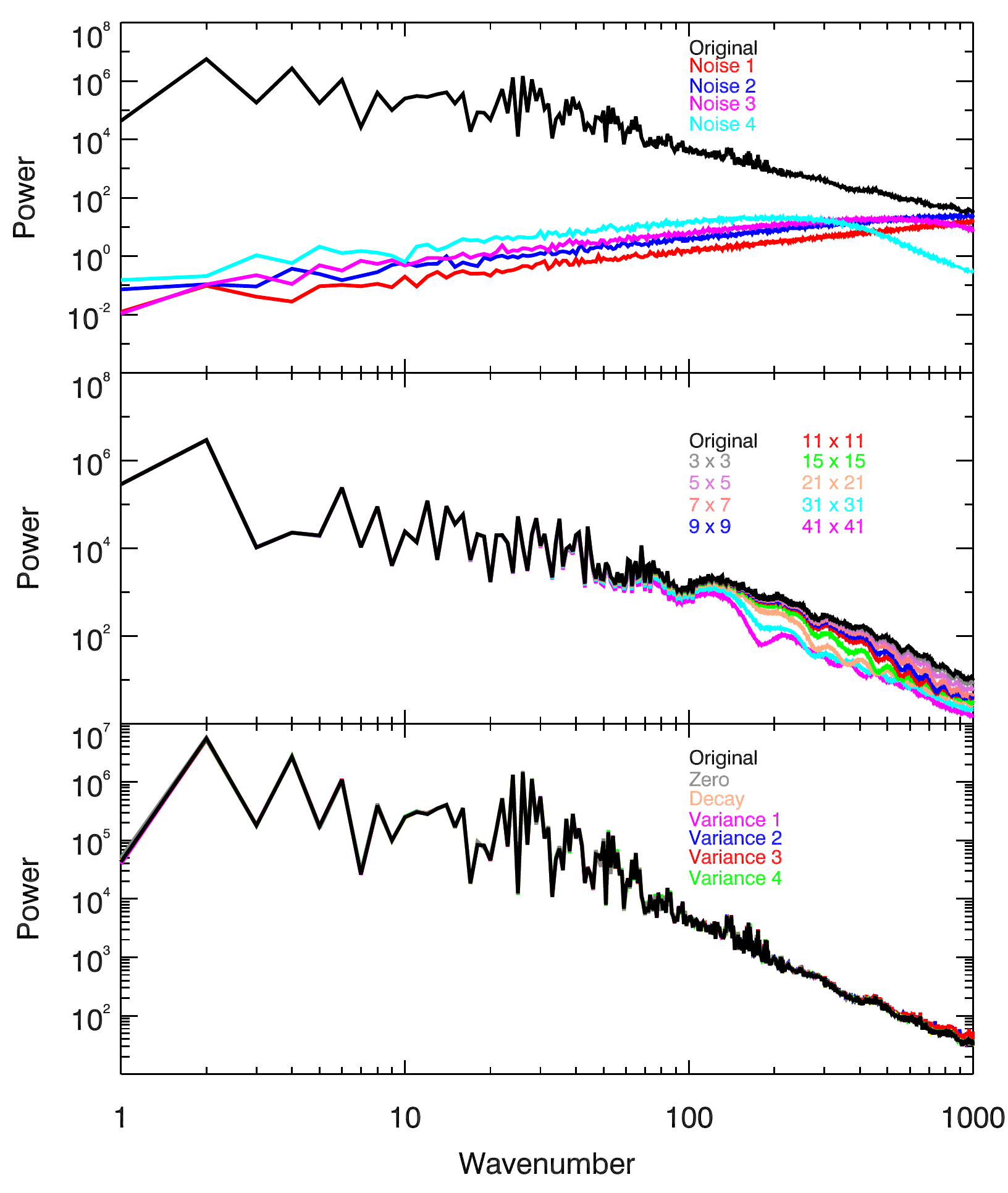}
  \caption[Test Figures]{
    \label{Figure: test_figs}
    \textbf{Top:} Comparison of the spectrum of Total Kinetic Energy Mosaic 1 with spectra of synthetic data possessing similar characteristics of measurement noise. Synthetic noise data sets \#1--4 have progressively longer length scales of 0.05$^{\circ}$, 0.15$^{\circ}$, 0.25$^{\circ}$, and 0.5$^{\circ}$. The typical rms amplitude of measurement noise in each synthetic data set is 15 m s$^{-1}$. \textbf{Middle:} Comparison of the spectrum of CB2 Mosaic 1 with spectra of the same mosaic with various smoothing filters applied to the image. The sizes of the smoothing boxes are listed in the figure for each test image, with the color used corresponding to each spectrum plotted in the figure. \textbf{Bottom:} Comparison of power spectra of kinetic energy maps derived from our original data and from test wind vector maps with different conditions for the high latitude winds. ``Zero'' and ``decay'' refer to the test cases with zero winds and a linear decay to zero winds at high latitudes, respectively. Variance 1 is the case with consistent variability throughout the high latitudes. Variances 2 and 3 are two distinct cases of variability applied to each row in the high latitudes from a random latitude in the original data set. Variance 4 is the ``mirror latitude'' case. 
    }
    \label{lastfig}
\end{figure}

Figure \ref{Figure: test_figs}a shows the power spectrum of one of our original kinetic energy mosaics along with spectra of synthetic noise data sets with various characteristic length scales. The rms values of the synthetic noise data in Figure \ref{Figure: test_figs}a are all $\sim$15 m s$^{-1}$, which is higher than expected noise values on average, but possible for certain isolated areas. The spectra of synthetic measurement noise in Figure \ref{Figure: test_figs}a have amplitudes that are several orders of magnitude below the spectrum of the kinetic energy map. Therefore, spectra of typical measurement noise have insufficient amplitude to adversely affect the calculated power spectrum of our data set. For measurement noise to influence the overall power spectrum, typical values of noise would need to be $\sim$40 m s$^{-1}$ or larger throughout the entire data set. We therefore conclude that measurement noise from our technique has little to no influence on the overall power spectrum.

\subsection{Smoothing}

To demonstrate the effect of data smoothing on the resulting power spectrum, we used one of our near-infrared CB2 image mosaics as the ``control'' case. We then smoothed the image with the size of the square smoothing box as a free parameter. Figure \ref{Figure: test_figs}b displays our results. Unsurprisingly, data smoothing dampens spectral power at high wavenumbers because variability at small length scales is diminished. However, spectral power at low wavenumbers is preserved, suggesting that this portion of the spectrum is relatively robust against smoothing. Regardless, we do not apply smoothing of our data at any point in our analysis, as we wish to avoid complexities with interpreting a spectrum whose structure is affected at high wavenumbers. 

\subsection{High Latitude Data}
\label{hld}

The kinetic energy spectra in Fig.~\ref{Figure: sph_powspec} use our full 2D wind vector maps at latitudes equatorward of $50^{\circ}$ but adopt winds that are zonally symmetric at latitudes higher than 50$^{\circ}$. In reality, of course, Jupiter's atmosphere at high latitudes contain numerous vortices and eddies; these features contribute spectral power at high wavenumbers that could affect the global power spectrum. Therefore, we investigated how close our original calculated spectrum should be to the true spectrum by testing other methods of completing the data set at high latitudes.

First, we tested cases where the high latitude winds are assumed to be zero everywhere or experience a linear decay to zero to the poles. (In Figure \ref{Figure: test_figs}c, these cases are labeled as ``Zero'' and ``Decay,'' respectively.) These cases retain zero longitudinal variability in the winds. Next, we tested cases of longitudinal variability at high latitudes using similar methodology. We extract a zonal strip of wind vector data from our results and subtract off the mean zonal-wind of the strip, resulting in the deviation from the mean as a function of longitude. We define the zonal variability as this deviation function, which we then add to the vectors at high latitudes. We then calculate the spectrum of the kinetic energy using the new wind map. In one case (Variance 1), we extracted a strip from each hemisphere near the 50$^{\circ}$ boundary of the original data, calculated the variability for each strip, and applied the variability throughout the high latitudes in the appropriate hemisphere. (This resulted in consistent variability throughout the high latitudes in each hemisphere.) In another scenario, we extracted a strip of data from a random latitude in the original data set and applied its variability to the high latitude data, repeating for each discrete row of high latitude data. We tested two separate cases (Variance 2 and Variance 3) with this scenario, each with distinct variabilities, for the high latitudes. Finally, we applied variability by using 50$^{\circ}$ as a ``mirror lattitude'' (Variance 4): winds at 55$^{\circ}$ latitude would have the same variability as winds at 45${\circ}$, 60$^{\circ}$ as 40$^{\circ}$, and so on.

Figure \ref{Figure: test_figs}c compares one of our total kinetic energy spectra with spectra calculated from these test cases. None of the spectra from our test cases are significantly different from our original spectrum with zero zonal variability in the high latitudes. The only perceptible difference in the spectra are for the Variance 2 and 3 cases. However, these scenarios overestimate the true spectral power present at the shortest wavelengths because their meridional variance is artificially high;  distinct variability is applied to each discrete latitude from random latitudes in the original wind vector maps. Overall, these sensitivity tests suggest that our spectrum may be fairly representative of the true atmospheric energy spectrum. Confirmation of this would require detailed wind maps of Jupiter's polar latitudes.   

\section{Discussion}

Our results indicate that 2D power spectra of both kinetic energy and cloud patterns resemble the predictions of 2D turbulence theory. This provides suggestive evidence that an inverse energy cascade is present within Jupiter's atmosphere, though the relative importance of isotropic and anisotropic inverse energy cascade are left for future analysis with 1D spectra from our data set. Our analysis further supports previous proposals that shallow forcing (mechanisms driving the flow occurring within the outer weather layer of Jupiter's atmosphere rather than in the deep convective interior) can drive and maintain the jet streams, as inverse cascade would be one possible mechanism of energy transfer. Recent numerical simulations have shown that forcing strictly confined to the weather layer can reproduce the bulk structure of Jupiter's and Saturn's jet streams, including the super-rotating equatorial jet \citep{Lian08, Lian10}. The exact nature of the shallow forcing is unknown, but one suggested mechanism is latent heat release from moist convection in the atmosphere \citep{Ingersoll00}, such as that observed in vigorous thunderstorms with embedded lightning \citep{Gierasch00}. Furthermore, observational studies \citep{Beebe80, Ingersoll81, Salyk06, DelGenio07} have all claimed to measure turbulent eddies presumably generated by the shallow forcing driving the jet streams at the cloud level on both Jupiter and Saturn. However, jet formation may not require an inverse cascade. Recent numerical simulations by \citet{Schneider09} successfully reproduce bulk characteristics of Jupiter's jet streams and equatorial superrotation without depending on an inverse energy cascade.  If this is the case, an inverse cascade may instead play a role in the transfer of energy from turbulent eddies to large-scale flows such as Jupiter's cyclones and anticyclones, rather than the jet streams. 

The classic interpretation for the transition in the power spectrum slopes is that it marks the scale at which a forcing mechanism supplies energy to Jupiter's atmosphere. Baroclinic instabilities and moist convection are proposed candidates for this forcing mechanism.  In either case, the transition provides information about the instability length scales and therefore the atmospheric stratification.  Baroclinic instabilities generally have peak growth rates close to the Rossby radius fo deformation, $L_d$ \citep{Pedlosky87}.   Similarly, studies suggest that discrete storms resulting from moist convection may also have scales similar to the deformation radius \citep{Lian10, Showman07}.  The radius of deformation is the length scale at which rotational or Coriolis effects become as significant as regular buoyancy effects in influencing the fluid flow and is defined as
\begin{equation}
L_d = \frac{NH}{f}
\end{equation}
where $N$ is the Brunt-V\"ais\"al\"a frequency, $H$ is the vertical scale of the flow (often taken as an atmospheric scale height), and $f$ is the Coriolis parameter. For the eddy kinetic energy power spectra, we measure the average transition wavenumber to be $\sim$135. This implies a transition wavelength of $\sim$3,000 km at a latitude of 45$^{\circ}$. The transition wavenumber for the 2D image mosaic spectra ranges from 200-300, corresponding to wavelengths $\sim$1,500--2,000 km, slightly lower than the estimate using the eddy kinetic energy power spectra.   Thus, our spectra tentatively suggest a Rossby deformation radius in the range 1000--$3000\,$km, similar to the estimates made by previous authors \citep[e.g.,][]{Cho01}.

Our present study also demonstrates that on Jupiter, typical cloud mosaic power spectra are distinct from kinetic energy power spectra. This result contrasts with previous studies that have demonstrated that the energy spectrum of an atmosphere and the power-law distribution of passive tracers (a constituent of an atmosphere that does not affect its flow, such as trace chemical concentrations and presumably chromophores on Jupiter that produce some cloud patterns) are similar. \citet{Nastrom86} examined the power-law spectra of the distribution of trace constituents (e.g. ozone, carbon monoxide, and water vapor) in Earth's atmosphere and found that the power spectrum approximately followed a -5/3 power-law distribution at scales less than 500--800 km. They further suggested that the distribution of trace chemical species in Earth's atmosphere is consistent with quasi-two-dimensional turbulence theory and is a direct result of the spectrum of displacements acting upon the tracers. A more recent study by \citet{Cho99} confirms the results shown by \citet{Nastrom86} with their measurements of ozone, methane, carbon monoxide, and carbon dioxide mixing ratios, all with spectral slopes ranging around -5/3. 

Turbulence theory provides insights into the expected slopes of passive-tracer spectra under idealized conditions. If we consider a simple dye (passive tracer) injected into a turbulent fluid, whose variance is created at a single, well-defined length scale, the spots of dye will slowly become elongated as quasi-random turbulent motions of the fluid interact with the dye patches, eventually producing a filamentary pattern. The length scale of the tracer variance will therefore change from large to small as the flow evolves, i.e., there will be a downscale cascade of tracer variance.  If this tracer variance is introduced at scales exceeding the energy forcing length scale, then at length scales in between the two forcing scales (i.e., in the inverse energy cascade regime) the tracer variance exhibits a spectrum with a spectral slope of -5/3 \citep{Batchelor59, Vallis06}.  On the other hand, at length scales smaller than both forcing scales---in the enstrophy cascade regime---the tracer-variance spectrum exhibits a slope of -1 \citep{Vallis06}.  This theory therefore suggests that the tracer variance spectrum exhibits a break in slope at the energy forcing scale but, unlike the kinetic energy spectrum, the predicted tracer-variance spectrum becomes {\it shallower} rather than {\it steeper} on the high-wavenumber side of this transition.  (The predicted slope will finally steepen again at very high wavenumbers where diffusion of the tracer variance becomes important.)  This behavior disagrees with our image-brightness spectra, which are steeper at high wavenumbers than low wavenumbers.

Why do our image brightness spectra differ from the predictions of turbulence theory? There are several possible reasons. First, cloud material (and therefore the variance of cloud brightness) on Jupiter is probably generated and destroyed over a broad range of horizontal scales, ranging from cumulus clouds created at the tens-of-km scale to stratus clouds created and destroyed at the belt/zone scale by meridional circulations. Moreover, the cloud brightness is affected not only by the column density of cloud material but by the particle size, single-scattering albedo, and chromophore abundance, and these properties may evolve in time over a broad range of spatial scales depending on the extant microphysical and chemical processes.  Creation and destruction of cloud variance may also be inhomogeneous (varying between belts and zones, for example) and anisotropic (with different spectral properties in the zonal and meridional directions).  Together, these factors imply that the sources and sinks of cloud brightness are far more complex---in their spatial, spectral, and temporal structure---than the idealized representations adopted in turbulence theory.  This may cause alterations to the spectral slopes relative to the predictions of turbulence theory.  Second, clouds are not passive tracers but actively affect the flow through their mass loading, the latent heating/cooling associated with their formation/sublimation, and their effects on atmospheric radiative heating rates.  It is perhaps then not altogether surprising that cloud spectra do not match simple turbulence predictions and that cloud and kinetic energy spectra differ.

Although our wind vector maps assumed no zonal variability in Jupiter's winds at latitudes higher than 50$^{\circ}$, our testing of various scenarios regarding the zonal variability of winds at these latitudes revealed that our spectrum is relatively robust. Confirmation of this would require detailed wind maps of Jupiter's atmosphere at high latitudes. Thus, our study illustrates the lack of knowledge regarding the dynamical meteorology of Jupiter at polar latitudes, and underscores the need for future spacecraft missions to address this deficiency in our observational record of the Solar System. Observations of the polar Jovian atmosphere with detailed imagery suitable for cloud tracking would illuminate the mesoscale structure of winds at these latitudes and advance our understanding of the overall atmosphere.

\section{Acknowledgements}
We thank Boris Galperin and an anonymous referee for helpful comments that strengthened this manuscript. Lorenzo Polvani, Peter Read, and Robert Scott provided valuable advice for this project. We thank Ashwin Vasavada for his work in compiling the raw \emph{Cassini} images and producing mosaics for the PDS Atmospheres node. This research was supported by a NASA Jupiter Data Analysis Program grant, \#NNX09AD98G, as well as a NASA Earth and Space Science Fellowship, \#NNX08AW01H. Additional support was provided by a University of Arizona TRIF Imaging Fellowship.

\label{lastpage}

\bibliographystyle{icarus} 
\bibliography{ICARUS_11709}

\end{document}